\documentstyle[12pt]{article}
\def\pd{\partial}
\def\be{\begin{equation}}
\def\ee{\end{equation}}
\def\bea{\begin{eqnarray}}
\def\eea{\end{eqnarray}}
\def\e{\rm {e}}

\def\ll{\cal L}
\def\a{ \alpha}

\def\Dslash{D\!\!\!\!/}
\def\r#1#2{\raisebox{.2ex}{$\displaystyle
\mathop{#1}^{{\scriptstyle #2}\rightarrow}$}}
\def\l#1#2{\raisebox{.2ex}{$\displaystyle
 \mathop{#1}^{\leftarrow {\scriptstyle #2}}$}}
\begin{document}
\begin{flushright}
{\bf hep-th/9707190}
 YITP-97-40\\
DTP-97-37  \end{flushright}
{\Large
\centerline{{\bf Moyal Brackets in M-Theory}}
\vskip 0.5cm
\centerline{David B. Fairlie\footnote{ On research leave
from the University of Durham, U.K. }}
\vskip 0.5cm

\centerline{Yukawa Institute for Theoretical Physics,}
\vskip 10pt

\centerline{Kyoto 606-01, Japan}
\vskip 10pt

\centerline{\phantom{.}{\sl David.Fairlie@durham.ac.uk}}}

\begin{abstract}
The infinite limit of Matrix Theory in 4 and 10 dimensions is described in
terms of Moyal Brackets. In those dimensions there exists a Bogomol'nyi bound
to the Euclideanized version of these equations, which  guarantees that
solutions of the first order equations also solve the second  order Matrix
Theory equations. A general construction of such solutions in terms of a
representation of the target space co-ordinates as non-local spinor bilinears,
which are generalisations of the standard Wigner functions on phase space, is
given.
\end {abstract}
\noindent \rule{7in}{0.1em}
\vskip 0.4cm

\section{Introduction}

 The purpose of the present paper is to clarify the meaning of
the large $N$ limit in M-theory and discuss the features of the formalism
peculiar to the
situation of 2 and 8 transverse dimensions.  When this limit is approached by
way of the Moyal Bracket formalism, the matrix M-theory takes on an aspect
resembling a generalisation of the Moyal formulation of Quantum Mechanics
\cite{moyal2}, in terms of Wigner phase space distributions.
 The target space co-ordinates in 4 and 10-dimensions will  be represented in
terms of  matrix elements between spinor wave functions as Wigner
distributions. \cite{wig} The non-local character of M-theory \cite{bank} is
perhaps reflected in this construction. The paper begins with a short review of
the usual interpretation of the large $N$ limit
before developing the interpretation in terms of Moyal Brackets. The essential
simplicity of  the theory in 4 and 10 dimensions resides in the fact that the
Euclidean form of the Lagrangian can be
written as a sum of squares of first order terms with positive relative signs,
whose vanishing implies the second order equations of motion, thus giving rise
to BPS states. These first order equations may be interpreted as a
generalisation of the Liouville equation for distribution functions. This is
the basis of the suggested analogy with Quantum Mechanics. This situation is
analysed in detail to give a general principle for the construction of
solutions to these equations. The  10-dimensional situation is complicated by
the presence of  equations of constraint. In the 4-dimensional case these
constraints are absent, and an alternative solution procedure is outlined.
\section{Setting the  Scene}
The Lagrangian describing  matrix models has been described some time ago
\cite{martyr}.
In recent months the paper of Banks et al for a matrix model of M-theory
\cite{banks} has stimulated the production of a profusion of papers elaborating
the issue \cite{profuse}. We take one such paper, \cite{dvv}  as a convenient
reference point.
 The Lagrangian to be used, taken
directly from \cite{dvv}
takes the form
of a two-dimensional ${\cal N}=8$
supersymmetric $U(N)$ Yang-Mills theory with the action
\be
\label{dvv}
S = \frac{1}{2\pi\alpha'} \int {\rm Tr}\left((D_\mu X)^2 + \theta^T \Dslash
\theta +
g_s^2 F_{\mu\nu}^2 -\frac{1}{ g_s^2}[X^i,X^j]^2 +
\frac{1}{ g_s} \theta^T\gamma_i [X^i,\theta]\right)d\sigma d\tau.
\ee
Here the 8 scalar fields $X^i$ are $N\times N$ hermitian matrices, as
are the 8 fermionic fields $\theta^\a_L$ and $\theta^{\dot\a}_R$.
The fields $X^i$, $\theta^\a$, $\theta^{\dot\a}$ transform respectively
as the ${\bf 8}_v$ vector, and ${\bf 8}_s$ and ${\bf 8}_c$ spinor
representations of the $SO(8)$ R-symmetry group of transversal rotations.
The two dimesional world sheet is a cylinder parametrised by co-ordinates
$\sigma,\ \tau$, with $0\leq\sigma\leq 2\pi$.
This article goes on to identify vacuum configurations of the theory in the
infra-red limit ($g_s\mapsto 0$) with matrices unitarily equivalent to diagonal
ones i.e.
\be
\label{diag2}
X^\mu =U^{-1}x^\mu U
\ee
where $x^\mu$ are diagonal matrices with entries $x_{ii}^\mu$. In the simplest
description of the model each of the eigenvalues $x_{ii},\ \theta^\a_{ii},\
\theta^{\dot\a}_{ii}$ is considered to represent a Green-Schwarz string in
the light cone gauge, and the entire model is described in this limit as a gas
of $N$ such strings. In a more general situation envisaged in (\ref{dvv}) as
$\sigma$, the parameter in the compactified direction increases from $0$ to
$2\pi$, instead of matching $x_{ii}(0)$ with $x_{ii}(2\pi)$ several subsets of
 the eigenvalues may become permuted and the eigenvalues return to their
 original values (for a given subset) only after $m_k$ turns round the compact
 direction. In this case the theory describes  $\sum N_k$ strings, where
 \be
 \label{stringno}
 \sum m_kN_k=N.
 \ee
 The authors of (\ref{dvv}) then proceed to argue that the bulk of the energy
 in the infinite momentum frame will be carried by long strings in the limit of
 very large $N$.
 There is, however, another way to consider the infinite
limit which ends with a theory formulated in terms of Poisson Brackets. This
theory possesses some unusual features; the bosonic sector possesses hidden
invariances due to Bars \cite{barbar} which apparently are lost in the finite
matrix version, and which represent the remnants of volume preserving
diffeomorphisms. Furthermore, when the transverse space is either 2 or
8-dimensional the first order (bosonic) equations of motion can be written
which provide a Bogomol'nyi bound on the action. These first order equations
furthermore, also possess the additional symmetry. It is to be expected that
something similar happens in the case of 4 transverse dimensions, but this case
is not analysed in the present article. Another approach to BPS configurations
can be found in \cite{matsuo}.

\section{Passage to the Infinite Limit}
The passage to the $N\rightarrow\infty$ limit can be conceived as taking place
in two steps; First of all the fields are described by functions of two
`phase space' variables $\alpha,\ \beta$ as well as $\sigma,\ \tau$ which
parametrise the world sheet instead of matrices, and
expressions of the form $\int {\rm Tr}[X^i,X^j]^2 d\sigma$ by
\be
\label{dvv1}
\int(\frac{1}{\lambda}\sin \{X^i,X^j\})^2d\alpha d\beta d\sigma
\ee
where $\sin \{X^i,X^j\}$ is the sine, or Moyal Bracket, \cite{moyal}, with
deformation parameter $\lambda$ which is defined as the imaginary part of the
star  product (*);
\be
X^\mu*X^\nu =\lim_{\stackrel{\alpha'\rightarrow \alpha}{\beta'\rightarrow
\beta}} \e^{i\lambda (\pd_\alpha\pd_\beta'-\pd_\alpha'\pd_\beta)}
X^\mu(\alpha,\beta,\sigma)X^\nu(\alpha',\beta',\sigma).
\label{star}
\ee
The point of this construction is that in the limiting points
$\displaystyle{\lambda\rightarrow \frac{2\pi}{N}}$ the Moyal brackets
\cite{ffz} reproduce the commutators of $N\times N$ matrices through the
association of the components $X^\mu_{mn}$ of a matrix $X^\mu$ with the
Fourier modes of a function $X^\mu(\alpha,\beta,\sigma)$, periodic in
$\alpha,\ \beta$. (Strictly speaking, $N$ should be odd for this association to
work.)
The fermionic terms in (\ref{dvv}) are replaced by the cosine bracket, i.e. the
real part of the star product.
Thus the action becomes
\bea
\label{dvv2}
S_{MB} &=& \frac{1}{2\pi\alpha'} \int \left((D_\mu X)^2 +
\cos\{\theta^T ,\Dslash \theta\} +
g_s^2{\rm Tr} F_{\mu\nu}^2\right)
d\alpha d\beta d\sigma  d\tau\nonumber\\
 &~&- \int \left((\frac{1}{\lambda g_s^2}\sin \{X^i,X^j\})^2 -
\frac{1}{ g_s}\cos\{
\theta^T\gamma_i,\frac{1}{\lambda}\sin\{X^i,\theta\}\}\right)
d\alpha d\beta d\sigma d\tau.
\eea
The final step involves letting $N$ grow indefinitely, or equivalently, letting
$\lambda\rightarrow 0$. The final form of the action $S$ is then expressed in
terms of ordinary Poisson Brackets;
\bea
\label{dvv3}
S_{PB} &=& \frac{1}{2\pi\alpha'} \int \left((D_\mu X)^2 +
\theta^T ,\Dslash \theta +
g_s^2{\rm Tr} F_{\mu\nu}^2 -(\frac{1}{ g_s^2} \{X^i,X^j\})^2\right)d\alpha
d\beta d\sigma d\tau\nonumber\\
&~& + \int \left( \frac{1}{ g_s} \theta^T\gamma_i,\{X^i,\theta\}\right)
d\alpha d\beta d\sigma  d\tau.
\eea
In this interpretation the theory describes a  3-membrane wrapped around
the compactified direction. In order to treat the system dynamically, the
longitudinal and timelike modes should be re-instated.

\section{Membranes in 8 transverse dimensions}
A remarkable feature of the situation with 8 transverse directions is that the
theory admits a class of solutions obtainable from a first order formulation,
thanks to the existence of a  a self-dual (antisymmetric) 4-tensor
$T_{\mu\nu\rho\sigma}$ in 10-dimensions which is an extension of that
discovered in 8-dimensions \cite{cor} and which is an analogue
of the 4-dimensional fully antisymmetric tensor
$\epsilon_{\mu\nu\rho\sigma}$. The equations quoted below are of similar form
to those exhibited as equations (63) in \cite{ckz}, but with a more physically
appropriate reinterpretation of the nature of the variables.
$X^0$ describes the timelike directions, $X^9$ the longitudinal and $X^i,\
i=1\dots 8$ the transverse directions. We shall also ignore the dependence upon
$\tau$, which is usually interpreted as world-sheet time and treat the
additional independent variable as $\sigma$ . This procedure is tanatamount to
working in a Euclidean, rather than Lorentzian space, as it is characteristic
of equations of self-duality that they should be formulated in a Euclidean
space to ensure that solutions will be real. This is the same sign  as that
adopted in \cite{becker}.

Then the 10-dimensional membrane
has a Lagrangian density in the
form of an $SU(\infty)$ pure Yang Mills theory (\ref{dvv3}). The situation is
further simplified to one of dependence  upon only one of the variables
$\sigma$ (apart from the $\alpha,\ \beta$ dependence of the gauge potentials
 $X^\mu(t,\alpha,\beta)$). We work in a gauge where $X^0=$ constant.
 The Lagrangian density is
 \be
 {\ll} = \frac{1}{2}(\pd_\sigma X^\mu)^2 +\frac{1}{4}\{X^\mu,\ X^\nu\}^2
 \label{bars1}
 \ee
 where curly brackets denote the Poisson Bracket with respect to the variables
 $\alpha,\ \beta$, i.e.
 \be
 \{X^\mu,\ X^\nu\}=\frac{\pd X^\mu}{\pd\alpha}\frac{\pd X^\nu}{\pd\beta}-
 \frac{\pd X^\mu}{\pd\beta}\frac{\pd X^\nu}{\pd\alpha}
 \label{bars2}
 \ee
 and the theory is equivalent to an SU$(\infty)$ Yang Mills with
  dependence upon only one variable. It is perhaps worth recalling that in the
case of finite $N$ Savvidy and others have
  shown that the classical second order equations for such a theory  in the
case when that variable is time
 exhibit chaotic solutions \cite{savvy}. It is not clear whether this statement
persists
when the large $N$ limit is taken.


The first  order set of equations is then given by
\bea
\partial_\sigma X^1+\{X^2,\ X^9\}&=&0\nonumber\\
\partial_\sigma X^2+\{X^9,\ X^1\}&=&0 \nonumber\\
\partial_\sigma X^3+\{X^4,\ X^9\}&=&0\nonumber\\
\partial_\sigma X^4+\{X^9,\ X^3\}&=&0\nonumber\\
\partial_\sigma X^5+\{X^6,\ X^9\}&=&0\nonumber\\
\partial_\sigma X^6+\{X^9,\ X^5\}&=&0\nonumber\\
\partial_\sigma X^7+\{X^8,\ X^9\}&=&0\nonumber\\
\partial_\sigma X^8+\{X^9,\ X^7\}&=&0\nonumber\\
\partial_\sigma X^9+\{X^1,\ X^2\}+\{X^3 ,\ X^4\}+ \{X^5,\ X^6\} +\{X^7,\ X^8\}
&=&0\nonumber\\
\{X^1,\ X^3\}+\{X^4 ,\ X^2\}+\{X^5,\ X^7\} +\{X^8,\ X^6\}&=&0\nonumber\\
\{X^1,\ X^4\}+\{X^2 ,\ X^3\}+\{X^8,\ X^5\} +\{X^7,\ X^6\}&=&0\nonumber\\
\{X^1,\ X^5\}+\{X^4 ,\ X^8\}+\{X^7,\ X^3\} +\{X^6,\ X^2\}&=&0\nonumber\\
\{X^1,\ X^6\}+\{X^2 ,\ X^5\}+\{X^3,\ X^8\} +\{X^4,\ X^7\}&=&0\nonumber\\
\{X^1,\ X^7\}+\{X^3 ,\ X^5\}+\{X^8,\ X^2\} +\{X^6,\ X^4\}&=&0\nonumber\\
\{X^1,\ X^8\}+\{X^5 ,\ X^4\}+\{X^2,\ X^7\} +\{X^6,\ X^3\}&=&0.
\label{nahm10}
\eea
It is easy to verify that  the second order equations coming from the
Lagrangian (\ref{bars1}) are satisfied by solutions of this system and that the
sums of squares of these equations give the Lagrangian density up to
divergences whether the brackets $\{\}$ are taken to signify commutators, Moyal
Brackets or Poisson brackets. Thus the solutions provide a bound on the
action. In the case of Poisson brackets, in fact only the cross terms
involving a $\sigma$ derivative give divergences; the others vanish
identically in virtue
of what we call the `Saturday Identity',
$$
\{f,\ g\}\{h,\ k\} + \{f,\ h\}\{k,\ g\}+\{f,\ k\}\{g,\ h\}\equiv 0,
\label{saturday}
$$
which holds for Poisson Brackets on a 2-dimensional phase-space-but not for
commutators, nor for the Moyal Bracket form. These equations will give ries to
solitonic solutions.

What we have in (\ref{nahm10}) in the case where $X^0=$ constant is a set of
coupled Nahm equations;
When expressed in terms of complex combinations defined by
\be
 z_1=X^1+iX^2;\ \   z_2=X^3+iX^4;\ \ z_3=X^5+iX^6;\ \ z_1=X^7+iX^8;\ \ H=X^9,
 \label{def}
 \ee
 these take the following form
 \bea
 \partial_\sigma z_1 + i\{H,\ z_1\}&=&0\nonumber\\
 \partial_\sigma z_2 + i\{H,\ z_2\}&=&0\nonumber\\
 \partial_\sigma z_3 + i\{H,\ z_3\}&=&0\nonumber\\
 \partial_\sigma z_4 + i\{H,\ z_4\}&=&0\nonumber\\
 \partial_\sigma H+\{\bar z_1,\ z_1\}+\{\bar z_2,\ z_2\}+\{\bar z_3,\ z_3\}+
 \{\bar z_4,\ z_4\}&=&0\nonumber\\
 \{z_1,\ z_2\}+\{\bar z_3,\ \bar z_4\}&=&0\nonumber\\
 \{z_1,\ z_3\}+\{\bar z_4,\ \bar z_2\}&=&0\nonumber\\
 \{z_1,\ z_4\}+\{\bar z_2,\ \bar z_3\}&=&0.
 \label{complexnahm}
 \eea
 \section{Steps towards a solution}
The main purpose of writing the equations in complex form is to emphasise
the similarity of the equations to the phase space formulation of Quantum
Mechanics.

 The observation  that the first four
  equations are of the (Quantum) Liouville type provides a clue towards a
  possible form of solution. To emphasise the similarity with Quantum
  Mechanics, the variables with respect to which the brackets are calculated
  are relabelled as $x,\ p$ instead of $\alpha,\ \beta$.
   Take the case where the bracket is
 Moyal; Then the work of Moyal showed that if $\psi(x,t)$ obeys the
Schr\"odinger equation
\be
i\hbar\frac{\pd}{\pd t}\psi =H\psi,
\label{sacred}
\ee
then the Wigner distribution function
\be
 f(x,p,t)=\int_{-\infty} ^\infty\bar\psi(x-y,t)\psi_k(x+y,t){\e}^{\frac{i py}
 {\hbar}}dy
 \label{wig}
 \ee
satisfies the quantum Liouville equation
\be
\frac{\pd}{\pd t}f(x,p,t)=\{H(x,p),f(x,p,t)\}_{MB},
\label{moy}
\ee
where the bracket is the Moyal bracket.
Baker\cite{baker} proved a significant converse theorem; that any real valued
distribution function satisfying (\ref{moy}) together with a normalisation
condition, could be expressed in terms of a Wigner distribution function
where the wave function  satisfies the Schr\"odinger equation (\ref{sacred})
\footnote{A more recent discussion of this view of quantum mechanics can be
found in\cite{dbf}.}\rlap.
Baker's work suggests that one way to satisfy the equations (\ref{complexnahm})
is to take  an ansatz of the form
 \be
 z_k=\int_{-\infty} ^\infty
\chi_k(x-y,\sigma)\psi_k(x+y,\sigma){\e}^{\frac{2i\pi py}{\lambda}}dy
 \label{ansatz}
 \ee
 provided  $\psi_k, \ \ k=1\dots 4$ satisfy the Schr\"odinger equation
 \be
 H(x,\frac{i\pd}{\pd x},\sigma)\psi_k(x,\sigma) =
\frac{i\lambda}{2\pi}\frac{\pd}{\pd \sigma}\psi_k(x,\sigma)
 \label{srod}
 \ee
 and $\chi_k(x,\sigma)$ satisfy the conjugate equation.
 This result follows from the observation that
 \bea
&&\frac{i}{2}(H*z_k-z_k*H)\nonumber\\
&=&\sum_j\sum_k(-\lambda)^{m+n}\frac{{\pd_x}^m}{m!}\frac{{\pd_p}^n}{n!}
H(x,p,\sigma)(-\pd_x)^n\int_{-\infty} ^\infty
y^m\chi_k(x-y,\sigma)\psi_k(x+y,\sigma){\e}^{\frac{2i\pi py}{\lambda}}dy
\nonumber\\
 &~&- H\leftrightarrow z_k\nonumber\\
 &=&\int_{-\infty} ^\infty
 H(x+y,p+i\pd_x)\chi_k(x-y,\sigma)\psi_k(x+y,\sigma){\e}^{\frac{2i\pi py}
 {\lambda}}dy - H\leftrightarrow z_k\nonumber\\
 &=&\int_{-\infty} ^\infty
 H(x+y,i\pd_{x+y})\chi_k(x-y,\sigma)\psi_k(x+y,\sigma){\e}^{\frac{2i\pi py}
 {\lambda}}dy \nonumber\\
 &~&-\int_{-\infty} ^\infty
  H(x-y,i\pd_{x-y})\chi_k(x-y,\sigma)\psi_k(x+y,\sigma){\e}^{\frac{2i\pi py}
 {\lambda}}dy.
 \label{obs}
 \eea

In a similar fashion the products $z_j*z_k$ can be calculated thanks to the
following Lemma\footnote{This was suggested by an unpublished result of Ian
Strachan}\rlap.
\bea
{\e}^{py}f(x)*{\e}^{py'}g(x)&=&
\sum_j\sum_k{\e}^{p(y+y')}(-\lambda)^{m+n}\frac{(y'\pd_x)^m}{m!}f(x)
\frac{(-y\pd_x)^n}{n!}g(x)\nonumber\\
&=&{\e}^{p(y+y')}f(x+y')g(x-y)
\label{id}
\eea
Then, by definition
\be
z_j*z_k=\int_{-\infty}^\infty\int_{-\infty}^\infty
\chi_j(x-y,\sigma)\psi_j(x+y,\sigma)
{\e}^{\frac{2i\pi py}{\lambda}}\chi_k(x-y',t)\psi_k(x+y',\sigma)
{\e}^{\frac{2i\pi py'}{\lambda}}dydy'
\label{zeqn}
\ee
This gives after integration over the variable $y-y'$
\be
z_j*z_k=N_{jk}
\int_{-\infty} ^\infty
\chi_j(x-y-y',\sigma)\psi_k(x+y+y',\sigma){\e}^{\frac{2i\pi p(y+y')}
{\lambda}}d(y+y'),
\label{id2}
\ee
where $N_{jk}$ is a ($\sigma$ dependent) normalisation factor;
\be
N_{jk}=\int_{-\infty}^\infty \chi_k(x-u,\sigma)\psi_j(x+u,\sigma)du.
\label{id3}
\ee
There remain the 6 = 3 + conjugate equations of constraint.
One obvious way in which these can be
satisfied is to choose $N_{jk}=n_j\delta_{jk}$. This then leaves one further
equation to be solved, namely,
\be
\partial_\sigma H+\{\bar z_1,\ z_1\}+\{\bar z_2,\ z_2\}+\{\bar z_3,\ z_3\}+
 \{\bar z_4,\ z_4\}=0.
\label{time}
\ee
This equation determines the $\sigma$ dependence of $H$. However this procedure
gives only a limited class of solutions, except in the 4-dimensional situation,
which will be analysed again later. In order to solve the full 10-dimensional
constraints in general, it is necessary to modify the ansatz to take  into
account
the group theoretical structure of the self-duality relations in 8-dimensions.
\section{Solution of 10-d equations }
Suppose $\gamma^{j},\ j=1\dots8$ are the gamma matrices in 8-dimensions, which
admit a real representation by $16\times16$ antisymmetric matrices \cite{cor}
and $\psi$ is a 16 component spinor, which will be later subject to algebraic
constraints. Take as ansatz
 \bea
 X^k&=&i\int_{-\infty} ^\infty
\bar\psi(x-y,\sigma)\gamma^k\psi(x+y,\sigma){\e}^{\frac{2i\pi py}{\lambda}}dy,\
k=1\dots8
\nonumber\\
X^9&=&i\int_{-\infty} ^\infty
\bar\psi(x-y,\sigma)\tilde\gamma^9\bar\psi(x+y,\sigma){\e}^{\frac{2i\pi
py}{\lambda}}dy,\ \alpha=9
 \label{ansatz2}
 \eea
and leave $x^9$ as an unspecified anti-Hermitean matrix for the moment. By
construction all $X^k,\ k=1\dots8$, are real.
There is nothing in this choice of which violates any symmetry principle;
we are already dealing with a broken symmetry situation.
Consider the six equations of constraint in (\ref{nahm10}), which are identical
to six of the seven equations of self-duality in 8-dimensions proposed in
\cite{cor}.
It is relatively easy to see that these last six  will be satisfied
automatically if $\psi$ is constrained to be of the form
\be
\psi=P_1P_2P_3\tilde\psi
\label{proj}
\ee
where $\tilde\psi$ is arbitrary and $P_1,\ P_2,\ P_3$ are three mutually
commuting projection operators given by
\bea
P_1&=&\frac{1}{4}(1+\gamma^1\gamma^3\gamma^5\gamma^7)(1+\gamma^2\gamma^4\gamma^6\gamma^8)
\nonumber\\
P_2&=&\frac{1}{4}(1+\gamma^1\gamma^4\gamma^8\gamma^5)(1+\gamma^2\gamma^3\gamma^7\gamma^6)
\\
P_3&=&\frac{1}{4}(1+\gamma^1\gamma^6\gamma^4\gamma^7)(1+\gamma^2\gamma^5\gamma^3\gamma^8).
\nonumber
\label{proj2}
\eea
This leaves the equations involving $\sigma$ dependence; the most conspicuous
among which is
\be
\partial_\sigma X^9=-(\{X^1,\ X^2\}+\{X^3 ,\ X^4\}+ \{X^5,\ X^6\} +\{X^7,\
X^8\}).
\label{outst}
\ee
 The other equations involving $\sigma$ derivatives and  (\ref{outst}) are then
solved with the ansatz;
\be
\tilde\gamma^9=-(\gamma^1\gamma^2+\gamma^3\gamma^4+\gamma^5\gamma^6+\gamma^7
\gamma^8),
\label{new}
\ee
if $\psi$ has an exponential $\sigma$ dependence.
It might be thought that this method would lead to a general construction of
a $\sigma$ independent solution, by choosing instead of (\ref{proj2})
\be
P_4=\frac{1}{4}(1+\gamma^1\gamma^6\gamma^2\gamma^5)(1+\gamma^4\gamma^7\gamma^3\gamma^8),
\label{alt}
\ee
and choosing as projection
\be
\psi=P_1P_2P_4\tilde\psi
\label{proj4}
\ee
Then not only do the equations of constraint vanish, but so does the right hand
side of (\ref{outst}) and we should have, in principle, a solution to the
8-dimensional self  dual equations. However, this is not so, as it may be
verified that the product $P_1P_2P_4\equiv 0$, i.e. the projection operator
vanishes!
By taking only two projection operators, $P_2$ and $P_3$ in the construction
it is found that only 5 of the 6 equations of constraint vanish. The
consistency
of the system may then be restored by introducing another co-ordinate $X^0$,
which plays a similar role to  $X^9$.
\section{Digression on the 4-d Moyal Nahm equations}
There is an alternative way to look at the 4-dimensional version of the first
order equations, which is closer in spirit to the initial discussion in which
the equations are regarded as an extension of Quantum Mechanics. This
relationship
is obscured in the 10-dimensional case on account of the equations of
constraint. Consider the 4-dimensional Moyal-Nahm equations, where again
$X^0$ is set to a constant
\bea
\partial_\sigma X^1+\{X^2,\ X^3\}&=&0\nonumber\\
\partial_\sigma X^2+\{X^3,\ X^1\}&=&0 \nonumber\\
\partial_\sigma X^3+\{X^1,\ X^2\}&=&0.
\label{3d}
\eea
and asume that the functions $\psi_j(x)$ belong to a real orthonormal basis
over the real line. Then take as ansatz
\be
X^i = \sqrt{(E_j-E_i)(E_i-E_k)}{\e}^{(E_j-E_k)\sigma}\int_{-\infty} ^\infty
\psi_j(x-y){\e}^{\frac{2i\pi py}{\lambda}}\psi_k(x+y)dy
 \label{cycle}
\ee
together with cyclic replacements. The $E_j$ are constants. Then on account of
the result (\ref{id2})
the  Moyal-Nahm equations are satisfied identically. This solution
can be extended in an obvious manner by separating the functions into three
classes according to the residue of the index mod 3;
\bea
X^0 &=& \sum_{j=0}^{j=\infty}\sqrt{(E_{3j+1}-E_{3j})(E_{3j}-E_{3j+2})}
{\e}^{(E_{3j+1}-E_{3j+2})\sigma}\int_{-\infty} ^\infty
\psi_{3j+1}(x-y){\e}^{\frac{2i\pi py}{\lambda}}\psi_{3j+2}(x+y)dy\nonumber\\
X^1 &=& \sum_{j=0}^{j=\infty}\sqrt{(E_{3j+2}-E_{3j+1})(E_{3j+1}-E_{3j})}
{\e}^{(E_{3j+2}-E_{3j})\sigma}\int_{-\infty} ^\infty
\psi_{3j+2}(x-y){\e}^{\frac{2i\pi py}{\lambda}}\psi_{3j}(x+y)dy\nonumber\\
X^2 &=& \sum_{j=0}^{j=\infty}\sqrt{(E_{3j}-E_{3j+2})(E_{3j+2}-E_{3j+1})}
{\e}^{(E_{3j}-E_{3j+1})\sigma}\int_{-\infty} ^\infty
\psi_{3j}(x-y){\e}^{\frac{2i\pi py}{\lambda}}\psi_{3j+1}(x+y)dy\nonumber
\label{tricycle}
\eea

The link with the above discussion on the analogy with the phase space
formulation of Quantum Mechanics is as follows;
When the correspondence
$\displaystyle{p\rightarrow\frac{i\lambda\pd_x}{2\pi}}$
is implemented
\bea
X^3\psi_1(x){\e}^{-E_1\sigma} &=& \sqrt{(E_1-E_3)(E_3-E_2)}{\e}^{-E_2\sigma}
\int_{-\infty} ^\infty\psi_1(x-y){\e}^{\r
{y\pd_x}{}}\psi_2(x+y)dy\psi_1(x)\nonumber\\
&=& \sqrt{(E_1-E_3)(E_3-E_2)}{\e}^{-E_2\sigma}
\int_{-\infty} ^\infty\psi_1(x-y)\psi_2(x)\psi_1(x-y)dy\nonumber\\
&=& \sqrt{(E_1-E_3)(E_3-E_2)}\psi_2(x){\e}^{-E_2\sigma}
\label{eigenvale}
\eea
on account of the orthogonality relations.
Similarly, exploiting the substitution $y\mapsto -y$ in the integral in
(\ref{cycle}) a similar equation for the adjoint appears;
 \bea
\psi_2(x){\e}^{E_2\sigma}\bar X_3 &=&
-\sqrt{(E_1-E_3)(E_3-E_2)}{\e}^{E_1\sigma}
\int_{-\infty} ^\infty\psi_2(x)\psi_1(x+y){\e}^{-\l {y\pd_x}{}}\psi_2(x-y)dy
\nonumber\\
&=& -\sqrt{(E_1-E_3)(E_3-E_2)}\psi_1(x){\e}^{E_1\sigma}
\int_{-\infty} ^\infty\psi_2(x-y)\psi_2(x-y))dy\nonumber\\
&=& -\sqrt{(E_1-E_3)(E_3-E_2)}\psi_1(x){\e}^{E_1\sigma}
\label{eigenvert}
\eea
This pair of equations can be combined into a Schr\"odinger like form
with $X_3$ playing the role of Hamiltonian.
\subsection{Back to Matrix Models}
The  treatment   of the Moyal version of these first order Matrix Theory
equations given here allows a neat return to the Matrix formulation in the
following manner;  define
\be
X^k = \sum_m\sum_n A^k_{mn}(t)\int_{-\infty} ^\infty
\psi_m(x-y)\psi_n(x+y){\e}^{\frac{2i\pi py}{\lambda}}dy
\label{annesatz}
\ee
where the coefficients $A^k_{mn}$ are $\sigma$ or $\tau$ dependent elements of
infinite-dimensional matrices labelled by the same index as $X^k$.
The use of the Lemma (\ref{id2})
transforms the Moyal version of equations (\ref{nahm10}) into the infinite
matrix version of the equations. This manipulation shows how closely the
infinite
matrix formulation is tied to the Moyal treatment.
\section{Conclusions}
This article has been written to examine the Moyal version of Matrix theory, in
the belief that this formulation is the most appropriate for the discussion of
the large $N$ limit and for the investigation of parallels with Quantum
Mechanics.
 The essence of the solution to the first order equations lies in the
representation of the co-ordinates in terms of matrix elements of a spinor
in 8-dimensions. The construction carries echoes of a similar representation
of the 10-dimensional string in terms of bilinears \cite{manogue}, but here
a noteworthy feature of the construction is that it is intrinsically non-local.
It would appear that instead of `magic, mystery, or membrane'
which have been suggested as a possible etymology \cite{wit}, the M in
M-Theory really stands for Moyal!
\section*{ Acknowledgements}
This work was initiated at CERN, Geneva and completed at the Yukakawa Institute
for Theoretical Physics, Kyoto. Both institutions are thanked for their
hospitality. The visit to Japan is funded  partly by the Royal Society of
London, and partly by JSPS, Kyoto.
I am also indebted to Cosmas Zachos for many helpful discussions which have
greatly  improved the presentation of this paper.

 \end{document}